\documentclass[aps,prc,twocolumn,showpacs,showkeys,nofootinbib,superscriptaddress]{revtex4-1}

\usepackage{dcolumn}
\usepackage{bm}
\usepackage{amsmath}    
\usepackage{amsfonts} 
\usepackage{amssymb}
\usepackage{mathptmx}      
\usepackage{graphicx}   
\usepackage{mathtools}
\usepackage{graphics}
\usepackage{multirow}
\usepackage{bibentry}
\usepackage{braket}
\usepackage{algorithm}
\usepackage{algorithmic}
\usepackage{listings}
\usepackage{setspace}
\usepackage{slashed}
\usepackage{color}
\usepackage{soul}

\usepackage{xfrac}

\def\ep{\epsilon}

\def\la{\langle}
\def\ra{\rangle}

\def\lam{\lambda}

\def\be{\begin{equation}}
\def\ee{\end{equation}}
\def\bea{\begin{eqnarray}}
\def\eea{\end{eqnarray}}

\def\Journal#1#2#3#4{{#1} {\bf#2}, #3 (#4)}

\def\PRD{{\rm Phys. Rev.} D}

\usepackage{hyperref}
\hypersetup{colorlinks=true,citecolor=blue,linkcolor=blue,urlcolor=blue}

\makeatletter
\providecommand \@ifxundefined [1]{%
 \@ifx{#1\undefined}
}%
\providecommand \@ifnum [1]{%
 \ifnum #1\expandafter \@firstoftwo
 \else \expandafter \@secondoftwo
 \fi
}%
\providecommand \@ifx [1]{%
 \ifx #1\expandafter \@firstoftwo
 \else \expandafter \@secondoftwo
 \fi
}%
\providecommand \href@noop [0]{\@secondoftwo}%
\providecommand \href [0]{\begingroup \@sanitize@url \@href}%
\providecommand \@href[1]{\@@startlink{#1}\@@href}%
\providecommand \@@href[1]{\endgroup#1\@@endlink}%
\providecommand \@sanitize@url [0]{\catcode `\\12\catcode `\$12\catcode
  `\&12\catcode `\#12\catcode `\^12\catcode `\_12\catcode `\%12\relax}%
\providecommand \@@startlink[1]{}%
\providecommand \@@endlink[0]{}%
\providecommand \url  [0]{\begingroup\@sanitize@url \@url }%
\providecommand \@url [1]{\endgroup\@href {#1}{\urlprefix }}%
\providecommand \urlprefix  [0]{URL }%
\providecommand \selectlanguage [0]{\@gobble}%
\providecommand \bibinfo  [0]{\@secondoftwo}%
\providecommand \bibfield  [0]{\@secondoftwo}%
\providecommand \BibitemShut  [1]{\csname bibitem#1\endcsname}%
\let\auto@bib@innerbib\@empty

\begin{document}
%
\title{Self-consistent light-front quark model analysis of $B\to D\ell\nu_\ell$ transition form factors}
\author{ Ho-Meoyng Choi\\
{\em Department of Physics, Teachers College, Kyungpook National University,
     Daegu, Korea 702-701}}
\begin{abstract}
We investigate the transition form factors $f_+(q^2)$ and $f_-(q^2)$ [or $f_0(q^2)$] for the exclusive
semileptonic $B\to D\ell\nu_\ell$ $(\ell=e, \mu,\tau)$ decays in the standard light-front quark model (LFQM) based on the
LF quantization.
The common belief is that 
while $f_+(q^2)$ can be obtained without involving any treacherous contributions
such as the zero mode and the instantaneous contribution, $f_-(q^2)$ receives those treacherous
contributions since it involves at least two components of the current, e.g. $(J^+, J^-)$ or $(J^+, {\bf J}_\perp)$.
Contrary to the common belief, we show  in the Drell-Yan ($q^+=0$) frame
that $f_-(q^2)$ obtained from  $(J^+, J^-)$ gives identical result to $f_-(q^2)$ obtained from $(J^+, {\bf J}_\perp)$
without involving such treacherous contributions in the standard LFQM. 
In our numerical calculations, we obtain the form factors and branching ratios for $B\to D\ell\nu_\ell$ $(\ell=e, \mu,\tau)$  
and compare with the experimental data as well as
other theoretical model predictions. Our results for ${\rm Br}(B\to D\ell\nu_\ell)$ show reasonable agreement with the experimental data except
for the semitauonic $B^0\to D^-\tau\nu_\tau$ decay.
The ratio ${\cal R}(D)=\frac{{\rm Br}(B\to D\tau\nu_\tau)}{{\rm Br}(B\to D\ell'\nu_{\ell'})}$ $(\ell'=e,\mu)$ 
is also estimated and compared with the experimental data as well as other theoretical predictions.
\end{abstract}

\maketitle
\section{Introduction}
The semileptonic $B\to D\ell\nu_\ell$ $(\ell=e,\mu, \tau)$ decays have attracted a lot of attention
in extracting the exclusive Cabibbo-Kobayashi-Maskawa (CKM) matrix element $|V_{cb}|$.
Especially, the substantial difference for the ratio
${\cal R}(D)={\rm Br}(B\to D\tau\nu_\tau)/{\rm Br}(B\to D\ell'\nu_{\ell'})$ $(\ell'=e,\mu)$ between the experimental data
and the standard model (SM) predictions generated a great excitement in 
testing the SM and  searching for new physics beyond the SM. The 
experimental data,
 ${\cal R}^{\rm exp}(D)=0.440(58)(42)$ measured from BaBar~\cite{BaB12,BaB13} and
${\cal R}^{\rm exp}(D)=0.375(64)(26)$ from Belle~\cite{Belle15},
have shown an excess over the standard model (SM) prediction
${\cal R}^{\rm SM}(D)=0.299(3)$~\cite{Amhis}. Many theoretical efforts have been made in resolving the issue of ${\cal R}(D)$ anomaly and  
searching for new physics beyond the SM~\cite{Lat1,Lat2, HQEFT1,Bigi,LCSR1}.

We note that the $B\to D\ell\nu_\ell$ decays involve two transition form factors (TFFs),
i.e. the vector form factor $f_+(q^2)$ and the scalar form factor $f_0(q^2)$.
The analysis of both TFFs $f_{+,0}(q^2)$  for $B\to D$ transitions can be found in various theoretical
approaches such as the lattice QCD (LQCD)~\cite{Lat1,Lat2},
the light-cone sum rule (LCSR)~\cite{LCSR0,LCSR1,LCSR2,LCSR3}, and the light-front quark model (LFQM)~\cite{Cheng04}.
While ${\rm Br}(B\to D\ell\nu_{\ell})$  for the light lepton decay modes $(\ell=e,\mu)$ needs only $f_{+}(q^2)$, 
${\rm Br}(B\to D\tau\nu_\tau)$ for the heavy $\tau$ decay mode receives contributions from both $f_{+}(q^2)$
and $f_0(q^2)$. The ratio ${\cal R}(D)$ is in particular quite sensitive to the scalar form factor.
This leads us to speculate that  the scalar contribution is the main source of  ${\cal R}(D)$ anomaly 
and thus the new physics effect beyond the SM.
However, since the predictions of $f_0(q^2)$ as well as $f_+(q^2)$
are quite different for different theoretical approaches within the SM,
it is very important to obtain the reliable and self-consistent results for the
TFFs before drawing any sound conclusion from the ${\cal R}(D)$ anomaly.

The purpose of this paper is to present the self-consistent descriptions of the $B\to D\ell\nu_\ell$ 
TFFs in the standard LFQM based on the LF quantization~\cite{SPP}.
There have been many previous LFQM analyses for the semileptonic decays between two pseudoscalar
mesons~\cite{SLF1,SLF2,CLF1,CLF2,CJ09}.
In fact, there are two main kinds of LFQM, i.e. the standard LFQM~\cite{SLF1,SLF2}
and the covariant LFQM~\cite{CLF1,CLF2,CJ09}. 
In the standard LFQM, the constituent quark and antiquark
in a bound state are required to be on-mass shells and the spin-orbit wave function is
obtained by the interaction-independent Melosh transformation~\cite{Melosh} from the ordinary equal-time static spin-orbit
wave function assigned by the quantum number $J^{PC}$.
The main characteristic of the standard LFQM is to use the sum of the LF energy of the constituent quark and antiquark for the
the meson mass in the spin-orbit wave function and any physical observable can be obtained directly in three-dimensional
LF momentum space using the more phenomenologically accessible LF wave function such as 
Gaussian radial wave function $\phi(x, {\bf k}_\perp)$.
However, as the standard LFQM itself is not amenable to analyze the zero-mode contribution, the covariant LFQM
using the manifestly covariant Bethe-Salpeter (BS) model with the multipole type $q{\bar q}$ vertex
was introduced~\cite{CLF1},  in which the constituents are off-mass shell.
While the covariant BS model used in~\cite{CLF1,CLF2,CJ09} allows one to analyze all the treacherous points such the zero 
modes and the off-mass shell instantaneous contributions in a systematic way,  it is less realistic than the standard LFQM.
Thus, in an effort to apply such treacherous points found in the covariant BS model to the standard LFQM, 
the effective replacement~\cite{CLF1,CLF2,CJ09} of
the LF vertex function $\chi(x,{\bf k}_\perp)$ obtained in the BS model with the more 
realistic Gaussian wave function $\phi(x,{\bf k}_\perp)$ in the standard LFQM has been made.

However, through the analysis of the vector meson decay constant together with the twist-2 and twist-3 distribution amplitudes (DAs) of the
vector meson~\cite{CJ14}, we found the correspondence relation between $\chi$ and $\phi$
proposed in~\cite{CLF1,CLF2,CJ09} encounters the self-consistency problem, e.g. the vector meson decay constants 
obtained in the standard LFQM were found to differ
for different sets of the LF current components and polarization states of the vector meson~\cite{CJ14}.  
We also resolved this self-consistency problem in the same work~\cite{CJ14} by imposing
the on-mass shell condition of the constituent quark and antiquark, i.e. replacement of the physical meson mass $M$ with the invariant mass $M_0$
in the integrand of formulas for the physical quantities, 
in addition to the original correspondence relation between $\chi$ and $\phi$.
The remarkable finding from our new self-consistent correspondence relations (i..e. $\chi\to\phi$ and $M\to M_0$) between
the two models [see, e.g. Eq. (49) in~\cite{CJ14}]
was that both zero-mode and instantaneous contributions appeared in the covariant BS model became absent in the standard LFQM with the LF
on-mass shell constituent quark  and antiquark degrees of freedom.
We then extended our self-consistent correspondence relations to analyze the decay amplitude related with twist-2 and twist-3 DAs
of pseudoscalar mesons~\cite{CJ15,CJ17} and observed the same conclusion drawn from~\cite{CJ14}.

In the previous analysis~\cite{CLF1,CLF2,CJ09} of the semileptonic decays between two pseudoscalar mesons using the covariant BS model,
the LF covariant calculations was made in the Drell-Yan-West ($q^+=q^0+q^3=0$) frame (i.e. $q^2=-{\bf q}^2_{\perp} <0$), which is advantageous
in that only the valence contributions are needed unless the zero-mode contributions exist.
The form factor $f_+(q^2)$ was obtained only from the plus component ($J^+$) of the weak current $J^\mu$ without encountering
the zero-mode contribution. One needs, however,  two different components of the
current to obtain the form factor $f_0(q^2)$ [or $f_-(q^2)]$, 
and  $J^+$ and ${\bf J}_{\perp}=(J_x, J_y)$  were 
used to obtain it in~\cite{CLF1,CLF2,CJ09}~\footnote{While the method of Jaus~\cite{CLF1} and ours~\cite{CJ09} in obtaining the form factors
are slightly different, the final results for $f_-(q^2)$ are the same with each other, i.e. 
$f_-(q^2)$ (see Eq. (4.3) in~\cite{CLF1} and Eq. (42) in~\cite{CJ09}) was obtained from using both $J^+$ and ${\bf J}_\perp$.}.
However, $f_-(q^2)$ obtained from ($J^+, {\bf J}_\perp$) in the covariant BS model receives
not only the instantaneous contribution but also the zero mode due to the ${\bf J}_\perp$ component.
Employing the effective method presented in~\cite{CLF1,CLF2,CJ09}  to express the zero-contribution as a convolution of the zero-mode operator
with the initial and final state LF vertex functions, the form factor $f_-(q^2)$ can also be expressed 
as the convolution form between the initial- and final-states LF vertex functions $\chi(x, {\bf k}_\perp)$ in the
valence sector. To obtain $f_+(q^2)$ and $f_-(q^2)$ in the more realistic standard LFQM, the authors in~\cite{CLF1,CLF2,CJ09}
use the only correspondence relation between $\chi$ and $\phi$ without imposing the on-mass shell condition (i.e. $M\to M_0$).

In the recent work in~\cite{QC18}, the authors 
investigated the self-consistency of the form factor $f_-(q^2)$ obtained from $(J^+, {\bf J}_\perp)$ by applying 
both the old correspondence $(\chi\to\phi)$ and our new correspondence ($\chi \to\phi$ and $M\to M_0$)
between the BS model and the standard LFQM. From their numerical calculations, the authors found 
from $f_-(q^2)$ in the standard LFQM that
the zero-mode contribution to $f_-(q^2)$ is sizable for the case of using only $(\chi\to\phi)$ relation but
vanishes when using  ($\chi \to\phi$ and $M\to M_0$) relations.
This result is very supportive to assert that our new correspondence relations are universally applicable even to the
weak transition form factors for a self-consistent  description of the standard LFQM. 
In order to assert that the form factor $f_-(q^2)$ is truly self-consistent, however,
it is essential to show that $f_-(q^2)$ obtained in the $q^+=0$ frame is independent of the components of the current,
i.e.  $f_-(q^2)$ obtained from $(J^+, J^-)$ is the same as the one obtained from $(J^+, {\bf J}_\perp)$.

In this work, we shall show that our new correspondence relations ($\chi \to\phi$ and $M\to M_0$)
guarantee the self-consistent description for the weak decay constant of a pseudoscalar meson and
the semileptonic decays between two pseudoscalar mesons in the standard LFQM.
To show this, we shall prove 
that (1) the decay constant $f_{\cal P}$ of a pseudoscalar meson (${\cal P}$) is independent of the components of the current, and (2) 
$f_-(q^2)$ obtained from $(J^+, J^-)$ is exactly the same as the one obtained 
from $(J^+, {\bf J}_\perp)$ in the $q^+=0$ frame.
Those findings again entail that the zero-mode contribution
as well as the instantaneous one appeared in the covariant BS model became absent 
in the standard LFQM.

Although we do not consider in this analysis, 
the  $q^+\neq 0$ frame  may be  used  to  compute  the  timelike process  such as this semileptonic decay
but  then  it  is  unavoidable  to  encounter  the particle-number-nonconserving Fock state (or nonvalence) contribution~\cite{CJ01}.
The main source of difficulty in the LFQM phenomenology is the lack of information on
the non-wave-function vertex~\cite{BCJ1,BCJ2} in the nonvalence diagram arising from the quark-antiquark pair creation/annihilation. 
This should contrast with the usual LF valence wave function. In principle, there is a systematic program
as was discussed in~\cite{BH98} to include the particle-number-nonconserving amplitude to take into
account the nonvalence contributions. However, the program requires to find all the higher Fock-state wave
functions while there has been relatively little progress in computing the basic wave functions of hadrons from
first principles. In the very recent analysis~\cite{Tang20} of the semileptonic $B_c\to\eta_c(J/\psi)$ decays in the framework of basis LF quantization,
the frame dependence of the TFFs between $q^+=0$ and $q^+\neq 0$ frames 
is discussed. The main reason for the frame dependence comes from the ignorance of the nonvalence contribution 
in the $q^+\neq 0$ frame and it is not even possible to show that the form factors are independent of the components of the current 
in the $q^+\neq 0$ frame unless the nonvalence contribution is correctly taken into account.
However, our main findings in the $q^+=0$ frame may be incorporated in the same $q^+=0$ frame calculations of Ref.~\cite{Tang20}.

The paper is organized as follows: In Sec.~\ref{sec:II},  we  briefly review the decay constant $f_{\cal P}$ 
of a pseudoscalar meson in an exactly solvable model based on the covariant BS model of ($3+1$) dimensional fermion field theory.
We then present our LF calculation of $f_{\cal P}$ in the BS model using both plus 
and minus  components of the current and discuss the treacherous points
such as the zero-mode contribution and the instantaneous one when the minus component of the current is used.
Linking the covariant BS model to the standard LFQM with our universal 
mapping  between the two models~\cite{CJ14,CJ15,CJ17}, we obtain $f_{\cal P}$ from both plus and minus components of the current
 in the standard LFQM. Our main finding is that while $f_{\cal P}$ obtained from
the minus component of the current in the covariant BS model receives both the zero mode and the instantaneous contributions, 
$f_{\cal P}$ obtained from the minus component of the current in the standard LFQM is free from such treacherous contributions
and gives identical result with the one obtained from the plus component of the current.
In Sec.~\ref{sec:III}, we obtain the transition form factors $f_{\pm}(q^2)$ in the standard LFQM using the same procedure discussed in
Sec.~\ref{sec:II}. Especially, we explicitly show that 
$f_-(q^2)$ obtained from $(J^+, J^-)$ is exactly the same as the one obtained 
from $(J^+, {\bf J}_\perp)$ in the $q^+=0$ frame. This finding again supports the universality of our correspondence relations between
the covariant BS model and the standard LFQM.
In Sec.~\ref{sec:IV}, we show our numerical results for the semileptonic $B\to D\ell\nu_\ell$ $(\ell=e,\mu, \tau)$ decays.
In the Appendix, the explicit forms of the standard LFQM results for $f_{\pm}(q^2)$ are presented.

\section{Decay Constant }
\label{sec:II}
\subsection{$f_{\cal P}$ in the covariant BS model}
In the solvable model, based on the covariant BS model of
($3+1$)-dimensional fermion field theory,
the decay constant  $f_{\cal P}$ of a pseudoscalar meson (${\cal P}$) 
with the four-momentum $P$ and mass $M$ as a $q{\bar q}$ bound state
is defined by the matrix element of the axial vector current 
\be\label{eq1}
\la 0|{\bar q}\gamma^\mu\gamma_5 q|{\cal P}(P)\ra
= if_{\cal P} P^\mu.
\ee

The matrix element $A^\mu \equiv \la 0|{\bar q}\gamma^\mu\gamma_5 q|{\cal P}(P)\ra$ is given in the one-loop
approximation as a momentum integral
\be\label{eq2}
A^\mu = N_c
\int\frac{d^4k}{(2\pi)^4} \frac{ H_{\cal P} S^\mu} {(p^2 -m^2_1 +i\varepsilon) (k^2 - m^2_{q}+i\varepsilon)},
\ee
where  $N_c$ is the number of colors and 
$p =P -k$ and $k$ are the internal momenta carried by the quark and antiquark propagators
of mass $m_1$ and $m_q$, respectively. The $q\bar{q}$ bound-state vertex function $H_{\cal P}$ of  a pseudoscalar meson
is taken as  multipole ansatz, i.e. $H_{\cal P}(p^2,k^2)=g/(p^2-\Lambda^2+i\epsilon)$  where $g$ and $\Lambda$ are
constant parameters in this manifestly covariant model.
The trace term is given by
\be\label{eq3}
S^\mu  = {\rm Tr}\left[\gamma^\mu\gamma_5\left(\slash \!\!\!\!\! p +m_1 \right) \gamma_5
 \left(-\slash \!\!\!\! k + m_q \right) \right].
\ee
Performing the LF calculation,  we take the reference frame where $P=(P^+, P^-, {\bf P}_\perp)=(P^+, M^2/P^+,{\bf 0}_\perp)$
and use the metric  convention $a\cdot b =\frac{1}{2} (a^+ b^- + a^- b^+) - {\bf a}_\perp\cdot {\bf b}_\perp$. 
We then obtain the identity $\not\!\!q = \not\!\!q_{\rm on}  +\frac{1}{2}  \gamma^+\Delta^-_q$,
where $\Delta^-_q = q^- - q^-_{\rm on}$ and 
the subscript (on) denotes the on-mass shell  quark momentum,
i.e., $p^2_{\rm on}=m^2_1$ and $k^2_{\rm on}=m^2_q$.
Using this identity,  one can separate the trace term into the on-shell propagating part
$S^{\mu}_{\rm on}$ and the off-mass shell instantaneous one $S^{\mu}_{\rm inst}$
as $S^\mu = S^\mu_{\rm on} + S^\mu_{\rm inst}$.

By the integration over $k^-$ in Eq.~(\ref{eq2}) and closing the contour in the lower half of the complex $k^-$ plane, one
picks up the residue at $k^-=k^-_{\rm on}$ in the region of $0<k^+<P^+$ (or $0<x<1$) where 
$x=\frac{p^+}{P^+}$ and $1-x = \frac{k^+}{P^+}$ are the LF longitudinal momentum fractions of the quark and antiquark. 
We denote the valence contribution to $A^\mu$ that is obtained by taking $k^-=k^-_{\rm on}$ in the region of
$0<x <1$ region as $[A^\mu]^{\rm LF}_{\rm val}$.
Then the Cauchy integration formula for the
$k^-$ integration in the valence region of Eq.~(\ref{eq2}) yields
 \be\label{eq4}
[A^\mu]^{\rm LFBS}_{\rm val}
= \frac{i N_c}{16 \pi^3}\int^1_0\frac{dx}{(1-x)}\int d^2{\bf k}_\perp
\chi(x,{\bf k}_\perp) S^\mu_{\rm val},
\ee
where $\chi(x,{\bf k}_\perp) =\frac{g}{x^2(M^2 -M^2_0)(M^2 -M^2_\Lambda)}$ is the LF quark-meson vertex function
and $M^2_{0(\Lambda)}= \frac{{\bf k}^2_\perp + m^2_1(\Lambda^2)}{x} + \frac{{\bf k}^2_\perp + m^2_q}{ 1-x}$.
The trace term in the valence contribution is given by
$S^\mu_{\rm val} = S^\mu_{\rm on} + S^\mu_{\rm inst}$, where
$S^\mu_{\rm on} = 4 (m_1 k^\mu_{\rm on} + m_q p^\mu_{\rm on})$ and
$S^\mu_{\rm inst} = 2 (m_1  \Delta^-_k +  m_q \Delta^-_{p}) g^{\mu +}$.
We note from  $S^\mu_{\rm inst}$ that the off-shell instantaneous contributions are nonzero for the minus component
of the current while they are absent for the plus or perpendicular components  of the current.

In our previous work~\cite{CJ14}, we check the LF covariance of $f_{\cal P}$ obtained from Eq.~(\ref{eq4})
using two different components (i.e. $\mu=+$ and $-$) of the current.
We found that while $f^{(+)}_{\cal P}$ obtained from $\mu=+$ is free from the zero mode,  
$f^{(-)}_{\cal P}$ obtained from $\mu=-$ receives the zero mode. 
We also identified the zero-mode operator corresponding to the zero-mode contribution to $f^{(-)}_{\cal P}$ (see Eq. (B9) in~\cite{CJ14}).
Since the LF calculations of $f_{\cal P}$ obtained from Eq.~(\ref{eq4})
were explicitly shown in~\cite{CJ14,CJ15}, we recapitulate the essential features of obtaining the full LF result of $f^{(-)}_{\cal P}$.
Then, we focus on the self-consistent standard LFQM analysis of $f_{\cal P}$ using 
our new correspondence relations (i.e. $\chi\to\phi$ and $M\to M_0$).

For $\mu=+$, the full result of $f_{\cal P}$ can be obtained only from the valence contribution with the 
on-mass shell quark propagating part, i.e.
$S^+_{\rm full} =S^+_{\rm val}= S^+_{\rm on}$.
The full solution of the decay constant obtained from $\mu=+$ is given by~\cite{CJ14,CJ15}
\be\label{eq5}
[f^{(+)}_{\cal P}]^{\rm LFBS}_{\rm full}
= \frac{N_c}{4\pi^3}\int^1_0\frac{dx}{(1-x)}\int d^2{\bf k}_\perp
\chi(x,{\bf k}_\perp) \frac{S^+_{\rm on}}{4P^+},
\ee
where $S^+_{\rm full}=S^+_{\rm on}=4P^+{\cal A}_1$ and ${\cal A}_1= (1-x)\; m_1 + x\;m_q$.

For $\mu=-$, the valence contribution 
to the trace term comes not only from the on-shell propagating part but also from the off-shell instantaneous one, i.e.
$S^-_{\rm val} = S^-_{\rm on} + S^-_{\rm inst}$. However, the valence contribution itself is not equal to the manifestly
covariant result (or equivalently $[f^{(+)}_{\cal P}]^{\rm LFBS}_{\rm full}$) since the minus component of the current receives the zero-mode
contribution as shown in~\cite{CJ14}.
In~\cite{CJ14}, we also found the zero-mode operator $S^{-}_{\rm Z.M.}$ corresponding to the zero-mode contribution at the trace level, i.e.
$S^-_{\rm Z.M.} = \frac{4}{P^+} (m_q - m_1) (-Z_2)$
with $Z_2 = x (M^2 - M^2_0)+ m^2_1 - m^2_q + (1-2x) M^2$. 
Adding $S^-_{\rm Z.M.}$ to $S^-_{\rm val}$, we found that
$S^-_{\rm full}= S^-_{\rm val} + S^-_{\rm Z.M.}= 4 P^-  {\cal A}_1$.

That is,  in this manifestly covariant BS model,  the full solution $[f^{(-)}_{\cal P}]_{\rm full}^{\rm LFBS}$ obtained from $\mu=-$ is completely equal to 
$[f^{(+)}_{\cal P}]_{\rm full}^{\rm LFBS}$ only if the zero-mode contribution is included in addition to the valence contribution.
We should note that while $[f^{(+)}_{\cal P}]_{\rm full}^{\rm LFBS}=[f^{(+)}_{\cal P}]_{\rm on}^{\rm LFBS}$, 
$[f^{(-)}_{\cal P}]_{\rm full}^{\rm LFBS}=[f^{(-)}_{\cal P}]_{\rm on}^{\rm LFBS}+[f^{(-)}_{\cal P}]_{\rm inst}^{\rm LFBS}+[f^{(-)}_{\cal P}]_{\rm Z.M.}^{\rm LFBS}$.

For the sake of comparison with $[f^{(+)}_{\cal P}]_{\rm on}^{\rm LFBS}$  and also for later use in the standard LFQM analysis,
we display the result of  $[f^{(-)}_{\cal P}]_{\rm on}^{\rm LFBS}$ obtained from Eq.~(\ref{eq4}) with only the on-mass propagating part, $S^\mu_{\rm val}=S^{-}_{\rm on}$,
as follows
\be\label{eq6}
[f^{(-)}_{\cal P}]_{\rm on}^{\rm LFBS}
= \frac{N_c}{4\pi^3}\int^1_0\frac{dx}{(1-x)}\int d^2{\bf k}_\perp
\chi(x,{\bf k}_\perp) 
\frac{P^+ S^-_{\rm on}}{4M^2},
\ee
where $S^-_{\rm on}=4(m_1 k^-_{\rm on} + m_q p^-_{\rm on})$ with
$k^{-}_{\rm on}=\frac{{\bf k}_\perp^2 + m_q^2}{(1-x)P^+}$ and 
$p^{-}_{\rm on}=\frac{{\bf k}_\perp^2 + m_1^2}{xP^+}$.

\subsection{$f_{\cal P}$ in the standard LFQM}
\label{sec:IIIb}
In the standard LFQM~\cite{SLF1,SLF2,Cheng97,KT,CJ98,CJ99,CJ99PLB,Choi07,CJ07},
the
wave function of a ground state pseudoscalar meson
as a $q\bar{q}$ bound state is given by
\be\label{eq7}
\Psi_{\lam{\bar\lam}}(x,{\bf k}_{\perp})
={\phi(x,{\bf k}_{\perp})\cal R}_{\lam{\bar\lam}}(x,{\bf k}_{\perp}),
\ee
where ${\cal R}_{\lam{\bar\lam}}$ is the spin-orbit wave function 
that is obtained by the interaction independent Melosh transformation from the ordinary
spin-orbit wave function assigned by the quantum number $J^{PC}$.
The covariant form of ${\cal R}_{\lam{\bar\lam}}$ with the definite spin $(S, S_z)=(0,0)$
constructed out of the LF helicity $\lam({\bar\lam})$ of a quark (antiquark)
is given by
\be\label{eq8}
{\cal R}_{\lam{\bar\lam}}
=\frac{\bar{u}_{\lam}(p_q)\gamma_5 v_{{\bar\lam}}( p_{\bar q})}
{\sqrt{2}[M^{2}_{0}-(m_1 -m_{q})^{2}]^{1/2}},
\ee
which satisfies the unitarity condition,
$\sum_{\lam{\bar\lam}}{\cal R}_{\lam{\bar\lam}}^{\dagger}{\cal R}_{\lam{\bar\lam}}=1$.
Its explicit  matrix form is given by
\be\label{eq9}
{\cal R}_{\lam{\bar\lam}}
=\frac{1}{\sqrt{2}\sqrt{{\bf k}^2_\perp+{\cal A}^2_1}}
\begin{pmatrix}
-k^L &  {\cal A}_1\\ -{\cal A}_1 & -k^R
\end{pmatrix},
\ee
where $k^{R} = k_x + i k_y$ and $k^{L} = k_x - i k_y$.

For the radial wave function $\phi$ in~Eq.~(\ref{eq7}), we use the Gaussian wave function 
\be\label{eq10}
\phi(x,{\bf k}_{\perp})=
\frac{4\pi^{3/4}}{\beta^{3/2}} \sqrt{\frac{\partial
k_z}{\partial x}} {\rm exp}(-{\vec k}^2/2\beta^2),
\ee
where $\vec{k}^2={\bf k}^2_\perp + k^2_z$ and $\beta$ is the variational parameter
fixed by the analysis of meson mass spectra~\cite{CJ09,CJ99,CJ99PLB,Choi07}.
The longitudinal component $k_z$ is defined by $k_z=(x-\frac{1}{2})M_0 +
\frac{(m^2_{q}-m^2_1)}{2M_0}$, and the Jacobian of the variable transformation
$\{x,{\bf k}_\perp\}\to {\vec k}=({\bf k}_\perp, k_z)$ is given by
$\frac{\partial k_z}{\partial x}
= \frac{M_0}{4 x (1-x)}[ 1-
(\frac{m^2_1 - m^2_q}{M^2_0})^2]$.
%
The normalization of our Gaussian radial wave function is then given by
\be\label{eq11}
\int^1_0 dx \int \frac{d^2{\bf k}_\perp}{16\pi^3}
|\phi(x,{\bf k}_{\perp})|^2=1.
\ee

Using the plus component of the current, the standard LFQM calculation of Eq.~(\ref{eq1})
is obtained by
\be\label{eq12}
[f^{(+)}_{\cal P}]^{\rm SLF}_{\rm on}= \frac{\sqrt{2N_c}}{{8\pi^3}}\int^1_0 dx \int d^2{\bf k}_\perp
\frac{\phi(x,{\bf k}_\perp)}{\sqrt{{\bf k}^2_\perp + {\cal A}^2_1}}
 \frac{S^+_{\rm on}}{4P^+}.
\ee
We should note that the main differences between the covariant BS model and the standard LFQM are attributed to the different 
spin structures of the $q{\bar q}$ system (i.e. off-shellness vs on-shellness) and the different meson-quark
vertex functions ($\chi$ vs $\phi$).  In other words, while the results of the covariant BS model allow the nonzero binding
energy $E_{\rm B.E.}=M^2 -M^2_0$, the SLF result is obtained from the condition of on-mass shell quark and antiquark (i.e. $M\to M_0$).

To find the exact correspondence between the covariant BS model and the standard LFQM, we first compare the physical quantities which are 
immune to the the treacherous points such as the zero modes or the instantaneous contributions in the BS model.
In the case of pseudoscalar meson decay constant,  since $f^{(+)}_{\cal P}$ obtained from the plus component of the current satisfies this prerequisite condition, 
one can find the following correspondence relation, 
$\sqrt{2N_c} \frac{ \chi(x,{\bf k}_\perp) } {1-x}
 \to \frac{ \phi(x,{\bf k}_\perp) } {\sqrt{ {\cal A}^2_1 + {\bf k}^2_\perp }}$, by comparing 
$[f^{(+)}_{\cal P}]^{\rm LFBS}_{\rm full}=[f^{(+)}_{\cal P}]^{\rm LFBS}_{\rm on}$  in Eq.~(\ref{eq5})
and $[f^{(+)}_{\cal P}]^{\rm SLF}_{\rm on}$ in Eq.~(\ref{eq12}).
In most previous LFQM analyses, this correspondence ($\chi$ vs $\phi$) has  also been used for the mapping of other physical observables
contaminated by the treacherous points.

In our previous analysis~\cite{CJ14,CJ15,CJ17}, we found that the correspondence relation including only LF vertex functions brings about
the self-consistency problem, i..e. the same physical quantity obtained from different components of the current and/or the polarization vectors
yields different results in the standard LFQM.
Our new correspondence relations between the two models to iron out the self-consistency problem is given by~\cite{CJ14,CJ15,CJ17}:
\be\label{eq13}
\sqrt{2N_c} \frac{ \chi(x,{\bf k}_\perp) } {1-x}
 \to \frac{ \phi(x,{\bf k}_\perp) } {\sqrt{ {\cal A}^2_1 + {\bf k}^2_\perp }}, \;\; M\to M_0,
 \ee
that is, the physical mass $M$ included in the integrand of the BS
amplitude should be replaced with the invariant mass $M_0$ since the results in the standard LFQM
are obtained from the requirement of all constituents being on their respective mass shell.
We should note that the correspondence in Eq.~(\ref{eq13}) between the covariant model and the LFQM has been
verified through our previous analyses of pseudoscalar~\cite{CJ15} and pseudotensor~\cite{CJ17} twist-3 DAs of a pseudoscalar meson
and  the chirality-even twist-2 and twist-3 DAs of a vector meson~\cite{CJ14}.

The virtue of Eq.~(\ref{eq13}) to restore the self-consistency of the standard LFQM is that 
one can apply Eq.~(\ref{eq13}) only to the on-mass shell contribution in the BS model to get the full result
in the standard LFQM. In other words, the treacherous points (i.e. zero mode and the instantaneous contribution)
appeared in the covariant BS model are absorbed into the LF on-mass shell constituent quark and antiquark contributions and the full result
in the standard LFQM is obtained only from the on-shell contribution  regardless of the components of the currents being used.
This remarkable feature also can be seen in this analysis of decay constant of pseudoscalar meson obtained from the ``$-$" component 
of the currents.
That is, applying Eq.~(\ref{eq13}) to $[f^{(-)}_{\cal P}]_{\rm on}^{\rm LFBS}$ given by Eq.~(\ref{eq6}), we obtain
the SLF result for the minus component of the current as follows
\be\label{eq14}
 [f^{(-)}_{\cal P}]^{\rm SLF}_{\rm on} = \frac{\sqrt{2N_c}}{{8\pi^3}}\int^1_0 dx \int d^2{\bf k}_\perp
\frac{\phi(x,{\bf k}_\perp)}{\sqrt{{\bf k}^2_\perp + {\cal A}^2_1}}
\frac{P^+ S^{-}_{\rm on}}{4M_0^2}.
\ee
We confirm numerically that $[f^{(-)}_{\cal P}]^{\rm SLF}_{\rm on}=[f^{(+)}_{\cal P}]^{\rm SLF}_{\rm on}$, which
contrasts with the covariant BS model calculation, in which $[f^{(-)}_{\cal P}]^{\rm LFBS}_{\rm on}\neq [f^{(+)}_{\cal P}]^{\rm LFBS}_{\rm on}$.
We also should note that our confirmation for $[f^{(-)}_{\cal P}]^{\rm SLF}_{\rm on}=[f^{(+)}_{\cal P}]^{\rm SLF}_{\rm on}$ is
independent of the form of the radial wave function, e.g. the power-law type wave function such as  $\phi\propto \sqrt{{\partial k_z}/\partial x}(1+ {\vec k}^2/\beta^2)^{-2}$ 
also shows $[f^{(-)}_{\cal P}]^{\rm SLF}_{\rm on}=[f^{(+)}_{\cal P}]^{\rm SLF}_{\rm on}$.

\section{ Semileptonic Decays between two pseudoscalar mesons}
\label{sec:III}
The transition form factors for the ${\cal P}(P_1)\to {\cal P}(P_2)\ell\nu_\ell$ semileptonic decays between two pseudoscalar mesons
 are given by 
\be\label{eq15}
\la P_2|V^\mu|P_{1}\ra =  f_{+}(q^2)(P_1 + P_2)^{\mu} +
f_-(q^2)q^\mu,
 \ee
where $q^\mu=(P_1-P_2)^\mu$ is the
four-momentum transfer to the lepton pair($\ell\nu_\ell$) and
$m^2_\ell\leq q^2\leq (M_1-M_2)^2$.  
The two form factors $f_{\pm}(q^2)$ also satisfy
\be\label{eq16}
f_0(q^2) = f_+(q^2) +
\frac{q^2}{M^2_1-M^2_2}f_-(q^2).
\ee

The matrix element
${\cal M}^\mu\equiv\la P_2|V^\mu|P_1\ra$ in the BS model is given by
 \be\label{eq21}
 {\cal M}^\mu = iN_c\int\frac{d^4k}{(2\pi)^4} \frac{H_{p_1}{\cal T}^\mu H_{p_2}} {N_{p_1} N_{k} N_{p_2}},
 \ee
where $N_{k} = k^2 - m^2_q + i\ep$ and  $N_{p_j} = p^2_{j} - m^2_j + i\ep$ with $p_j=P_j -k\;(j=1,2)$. 
To be consistent with the analysis of the decay constant, we take
the $q\bar{q}$ bound-state vertex functions $H_{p_j}(p^2_j,k^2)=g_j/(p^2_j -\Lambda^2_j+i\epsilon)$
of the initial ($j=1$) and final ($j=2$) state pseudoscalar mesons.
The trace term is given by
\be\label{eq22}
{\cal T}^\mu  = {\rm Tr}[\gamma_5\left(\slash \!\!\!\!\! p_1 +m_1 \right) \gamma^\mu \left(\slash \!\!\!\!\! p_2 +m_2 \right)\gamma_5
 \left(-\slash \!\!\!\! k + m_q \right)].
 \ee

Performing the LF calculation of Eq.~(\ref{eq21}) in the valence region ($0<k^+<P^+_2$) of the $q^+=0$ frame, where the pole $k^-=k^-_{\rm on}=({\bf
k}^2_\perp + m^2_q  -i\ep)/k^+$ (i.e., the spectator quark) is located in the lower half of
the complex $k^-$ plane,  the Cauchy integration formula for the
$k^-$ integral in Eq.~(\ref{eq21}) gives
  \be\label{eq23}
 [{\cal M}^\mu]^{\rm LFBS}_{\rm val} =
N_c\int^1_0 \frac{dx}{(1-x)}\int \frac{d^2{\bf k}_\perp}{16\pi^3}
\chi_1(x,{\bf k}_\perp) \chi_2 (x, {\bf k'}_\perp)
 {\cal T}^\mu_{\rm val} ,
  \ee
where 
\be\label{eq24}
\chi_{1(2)} = \frac{g_{1(2)}}{x^2 (M^2_{1(2)} - M^{(\prime)2}_0)(M^2_{1(2)} -M^2_{\Lambda_{1(2)}})},
\ee
with
$M^2_{0(\Lambda_1)}= \frac{{\bf k}^2_\perp + m^2_1(\Lambda^2_1)}{x} + \frac{{\bf k}^2_\perp + m^2_q}{ 1-x}$
and $M'^2_{0(\Lambda_2)}=M^2_{0(\Lambda_1)}(m_1(\Lambda_1)\to m_2(\Lambda_2),  {\bf k}_\perp\to {\bf k'}_\perp={\bf k}_\perp + (1-x) {\bf q}_\perp$).
The explicit LF calculation of Eq.~(\ref{eq23}) in parallel with the manifestly covariant calculation of Eq.~(\ref{eq21}) can be found in~\cite{CJ09}.
As shown in Ref.~\cite{CJ09}, while $f_{+}(q^2)$ was obtained from $J^+$ and immune to the zero mode, 
the form factor $f_-(q^2)$ was obtained from $(J^+, {\bf J}_\perp)$ and received both the instantaneous and the zero-mode
contributions. Of course, one cannot avoid such treacherous points in the BS model
even if  $f_-(q^2)$ is obtained from the two components $(J^+, J^-)$ of the current.

In this work, we shall show that $f_-(q^2)$ in the standard LFQM is independent of the components of the current, i.e. regardless of using
$(J^+, {\bf J}_\perp)$ or $(J^+, J^-)$, as far as we apply Eq.~(\ref{eq13}) in the BS model to get the standard LFQM results.
So, from now on, we discuss only for the on-mass shell contribution in the valence region of the $q^+=0$ frame.
Of the trace terms ${\cal T}^\mu_{\rm val}= {\cal T}^\mu_{\rm on} + {\cal T}^\mu_{\rm inst}$, the on-shell
contribution is given by
\bea\label{eq25}
{\cal T}^\mu_{\rm on} &=&
4 \biggl[
 p^\mu_{1\rm on} (p_{2\rm on}\cdot k_{\rm on}) - k^\mu_{\rm on} (p_{1\rm on}\cdot p_{2\rm on})
+  p^\mu_{2\rm on} (p_{1\rm on}\cdot k_{\rm on})
\nonumber\\
&&\hspace{0.5cm}
+ m_2m_{\bar q} p^\mu_{1\rm on}  
 +  m_1m_{\bar q}p^\mu_{2\rm on}
+  m_1m_2 k^\mu_{\rm on} \biggr],
 \eea
 where 
\bea\label{eq26}
p_{1\rm on}&=& \left[ x P^+_1, \frac{m^2_1 + {\bf k}_\perp^2}{xP^+_1}, -{\bf k}_\perp \right],
\nonumber\\
p_{2\rm on} &=& \left[ x P^+_1, \frac{m^2_2 + ({\bf k}_\perp+{\bf q}_\perp)^2}{xP^+_1}, -{\bf k}_\perp-{\bf q}_\perp \right],
\nonumber\\
k_{\rm on} &=& \left[ (1-x) P^+_1, \frac{m^2_q + {\bf k}_\perp^2}{ (1-x)P^+_1}, {\bf k}_\perp \right].
\eea
The explicit form of the instantaneous contribution ${\cal T}^\mu_{\rm inst}$ can be found in~\cite{CJ09}.
On the one hand,  the transition form factors  $f_{\pm}(q^2)$ obtained from 
$(J^+, {\bf J}_\perp)$ are given by
 \bea\label{eq27}
 f_+(q^2) &=& \frac{{\cal M}^+}{2P^+_1},
\nonumber\\
 f^{(\perp)}_-(q^2) &=&  \frac{{\cal M}^+}{2P^+_1}+ \frac{ {\cal M}^\perp \cdot {\bf q}_\perp}{ {\bf q}^2_\perp}.
  \eea
 On the other hand, the form factor $f_-(q^2)$ obtained from $(J^+, J^-)$ is given by
 \be\label{eq28}
 f^{(-)}_-(q^2) = -  \frac{{\cal M}^+}{2P^+_1} \biggl( \frac{\Delta M^2_{+}  + {\bf q}^2_\perp}{\Delta M^2_{-}  - {\bf q}^2_\perp} \biggr)
  + \frac{ P^+_1 {\cal M}^-}{ \Delta M^2_{-}  - {\bf q}^2_\perp},
 \ee
where $\Delta M^2_{\pm} = M^2_1 \pm M^2_2$.  For convenience sake,  the form factor $f_-(q^2)$ obtained from $(J^+, {\bf J}_\perp)$ and $(J^+, J^-)$
is denoted by $f^{(\perp)}_-(q^2)$ and $ f^{(-)}_-(q^2)$, respectively.
 In the manifestly covariant BS model given by Eq.~(\ref{eq21}), we note that
while $[f^{(+)}_{+}]_{\rm full}^{\rm LFBS}=[f^{(+)}_{+}]_{\rm on}^{\rm LFBS}$,
$[f^{(\perp)}_{-}]_{\rm full}^{\rm LFBS}=[f^{(\perp)}_{-}]_{\rm on}^{\rm LFBS}+[f^{(\perp)}_{-}]_{\rm inst}^{\rm LFBS}+[f^{(\perp)}_{-}]_{\rm Z.M.}^{\rm LFBS}$.
The full result $f^{(-)}_-(q^2)$ has the same structure as $f^{(\perp)}_-(q^2)$,
i.e.
$[f^{(-)}_{-}]_{\rm full}^{\rm LFBS}=[f^{(-)}_{-}]_{\rm on}^{\rm LFBS}+[f^{(-)}_{-}]_{\rm inst}^{\rm LFBS}+[f^{(-)}_{-}]_{\rm Z.M.}^{\rm LFBS}$
although the explicit forms of the instantaneous and zero-mode contributions are different from those for $f^{(\perp)}_-(q^2)$.

For the calculation of the transition form factors $f_{\pm}(q^2)$, our new correspondence relations between the covariant BS model and the
standard LFQM is given by
\bea\label{eq29}
&&\sqrt{2N_c} \frac{ \chi_1(x,{\bf k}_\perp) } {1-x}
 \to \frac{ \phi_1(x,{\bf k}_\perp) } {\sqrt{ {\cal A}^2_1 + {\bf k}^2_\perp }}, \;\; M_1\to M_0,
 \nonumber\\
&& \sqrt{2N_c} \frac{ \chi_2(x,{\bf k'}_\perp) } {1-x}
 \to \frac{ \phi_2(x,{\bf k'}_\perp) } {\sqrt{ {\cal A}^2_2 + {\bf k'}^2_\perp }}, \;\; M_2\to M'_0.
 \eea

In order to obtain the self-consistent description of our standard LFQM,
we first compute $[f_{+}]_{\rm full}^{\rm LFBS}=[f_{+}]_{\rm on}^{\rm LFBS}$, $[f^{(\perp)}_{-}]_{\rm on}^{\rm LFBS}$, and $[f^{(-)}_{-}]_{\rm on}^{\rm LFBS}$ 
from the BS model and apply Eq.~(\ref{eq29}) to get the corresponding standard LFQM results, 
i.e. $[f_{+}]_{\rm on}^{\rm SLF}$, $[f^{(\perp)}_{-}]_{\rm on}^{\rm SLF}$ and $[f^{(-)}_{-}]_{\rm on}^{\rm SLF}$, respectively.
The final standard LFQM results for $f_{\pm}(q^2)$  are given by
\bea\label{eq30}
[ f_{+}(q^2)]^{\rm SLF}_{\rm on} &=& \int^{1}_{0}dx \int \frac{d^{2}{\bf k}_{\perp}}{16\pi^3}
\frac{\phi_{1}(x,{\bf k}_{\perp})}{\sqrt{ {\cal A}_{1}^{2} + {\bf k}^{2}_{\perp}}}
\frac{\phi_{2}(x,{\bf k}'_{\perp})}{\sqrt{ {\cal A}_{2}^{2}+ {\bf k}^{\prime 2}_{\perp}}}
\nonumber\\
&&\times 
\frac{(1-x)}{2} \left[\frac{ {\cal T}^{+}_{\rm on}} {2P^+_1}\right],
\eea
\bea\label{eq31}
[ f^{(\perp)}_{-}(q^2)]^{\rm SLF}_{\rm on} &=& \int^{1}_{0}dx \int \frac{d^{2}{\bf k}_{\perp}}{16\pi^3}
\frac{\phi_{1}(x,{\bf k}_{\perp})}{\sqrt{ {\cal A}_{1}^{2} + {\bf k}^{2}_{\perp}}}
\frac{\phi_{2}(x,{\bf k}'_{\perp})}{\sqrt{ {\cal A}_{2}^{2}+ {\bf k}^{\prime 2}_{\perp}}}
\nonumber\\
&&\times 
\frac{(1-x)}{2} \left[\frac{ {\cal T}^{+}_{\rm on}} {2P^+_1} + \frac{{\bf{\cal T}}_{\perp\rm on}\cdot{\bf q}_{\perp}}{{\bf q}^2_\perp}\right],
\eea
and
\bea\label{eq32}
[ f^{(-)}_{-}(q^2)]^{\rm SLF}_{\rm on} &=& \int^{1}_{0}dx \int \frac{d^{2}{\bf k}_{\perp}}{16\pi^3}
\frac{\phi_{1}(x,{\bf k}_{\perp})}{\sqrt{ {\cal A}_{1}^{2} + {\bf k}^{2}_{\perp}}}
\frac{\phi_{2}(x,{\bf k}'_{\perp})}{\sqrt{ {\cal A}_{2}^{2}+ {\bf k}^{\prime 2}_{\perp}}}
\nonumber\\
&&\times 
\frac{(1-x)[P^+_1{\cal T}^-_{\rm on}-\frac{ {\cal T}^{+}_{\rm on}} {2P^+_1} ({\Delta M}^2_{0+}+{\bf q}^2_\perp)]}
{2({\Delta M}^2_{0-}-{\bf q}^2_\perp)},
\nonumber\\
\eea
respectively, where ${\Delta M}^2_{0\pm} = M^2_0 \pm M^{\prime 2}_0$ 
obtained from the on-mass shell condition (i.e. $M^{(\prime)}\to M^{(\prime)}_0$)
and ${\cal A}_i  =  (1-x) m_i + x m_{q}\; (i=1,2)$. 
We numerically confirm that $[ f^{(\perp)}_{-}(q^2)]^{\rm SLF}_{\rm on}=[ f^{(-)}_{-}(q^2)]^{\rm SLF}_{\rm on}$,
which supports the self-consistency of our standard LFQM.
The explicit forms of the on-shell trace terms and the form factors
in Eqs.~(\ref{eq30})-(\ref{eq32}) are given in the Appendix.
We note that the form factors obtained in the spacelike region using
 the $q^+=0$ frame are analytically continued to the timelike region by changing ${\bf q}^2_\perp$
 to $-q^2$ in the form factors.

Including the nonzero lepton mass ($m_\ell$), the differential decay rate for the
exclusive ${\cal P}(P_1)\to {\cal P}(P_2)\ell\nu_\ell$ process is given by~\cite{KS, Yao}
\be\label{eq17} 
\frac{d\Gamma}{dq^{2}}=\frac{8 {\cal N} |{\vec p}^*|}{3}
\left[
\biggl(1+\frac{m^2_\ell}{2q^2}\biggr) |H_+|^{2}
+
\frac{3 m^2_\ell}{2 q^2}|H_0|^2
\right],
\ee
where 
\be\label{eq18} 
|{\vec p}^*| = \frac{1}{2M_{1}}\sqrt{
(M_{1}^{2}+M_{2}^{2}-q^{2})^{2}-4M_{1}^{2}M_{2}^{2}}
 \ee
 is the modulus of the three-momentum of the daughter meson in the parent meson rest frame and the helicity amplitudes
 $H_0$ and $H_t$ corresponding to the longitudinal parts of the spin-1 and spin-0 hadronic contributions, respectively,
 can be expressed in terms of  $f_+$ and $f_0$ as follows
 \be\label{eq19}
 H_+ = \frac{2 M_1 |{\vec p}^*| }{\sqrt{q^2}} f_{+}(q^2),\;\;
  H_0 = \frac{M^2_1- M^2_2 }{\sqrt{q^2}} f_{0}(q^2).
 \ee
The normalization factor in Eq.~(\ref{eq17}) is 
\be\label{eq20}
{\cal N} = \frac{G^{2}_{F}}{256\pi^{3}} \eta^2_{\rm EW} |V_{Q_{1}\bar{Q}_{2}}|^{2}\frac{q^2}{M^2_1}
\biggl(1-\frac{m^2_\ell}{q^2}\biggr)^2,
\ee
where $G_{F}=1.166\times 10^{-5}$ GeV$^{-2}$ is the Fermi constant, $V_{Q_{1}\bar{Q}_{2}}$ is the relevant CKM
mixing matrix element and the factor $\eta_{\rm EW} =1.0066$ accounts for the leading order electroweak corrections~\cite{Sirlin}.

The kinematics of the ${\cal P}(P_1)\to {\cal P}(P_2)\ell\nu_\ell$ decay can also be expressed in terms of
the recoil variable $w$ defined by
\be\label{rw}
w = v_{1}\cdot v_{2} = \frac{ M^2_1 + M^2_2 - q^2}{2 M_1 M_2},
\ee
where $v_{1(2)}=\frac{P_{1(2)}}{M_{1(2)}}$ is the four velocity of the initial (final) meson and $q^2=(P_1-P_2)^2=(P_\ell + P_\nu)^2$.
While the minimum value of $w=1$ (or $q^2=q^2_{\rm max}$) 
corresponds to zero-recoil of the final meson in the initial meson rest frame,
the maximum value of $w$ (or $q^2 =0$) corresponds to the maximum recoil of the final
meson recoiling with the maximum three momentum $|{\vec P}_2|=\frac{(M^2_1 - M^2_2)}{2 M_1}$.

\section{Numerical Results}
\label{sec:IV}
In our numerical calculations for the semileptonic
$B\to D\ell\nu_\ell$ ($\ell=e,\mu,\tau$) decays, we use two sets of
model parameters ($m,\beta$) for the linear and harmonic oscillator (HO)
confining potentials given in Table~\ref{t1} obtained from the calculation of the ground state
meson mass spectra~\cite{Choi07,CJ09}. For the physical $(B, D)$ meson masses, we
use the central values quoted by the Particle Data Group (PDG)~\cite{PDG}.
Our predictions for the decay constants of $(D, B)$ mesons obtained from the model parameters
in Table~\ref{t1} 
are $f_{D}=197\; (180)$ MeV  and $f_{B}=171\;(161)$ MeV
for the linear (HO) parameters, respectively, while the current available
experimental data are given by
$f^{\rm exp}_{D}=205.8(4.5)(0.4)(2.7)$ MeV~\cite{PDG}
and 
$f^{\rm exp}_{B}=229^{+39+34}_{-31-37}$ MeV~\cite{Ikado}.

\begin{table}
\centering
\caption{The constituent quark mass $m_q$ (in GeV) and the gaussian parameters
$\beta_{q{\bar q}}$ (in GeV) for the linear and HO confining potential
obtained by the variational principle~\cite{CJ09,Choi07}. 
$q=u$ and $d$.}
\label{t1}
\renewcommand{\arraystretch}{1.2}
\setlength{\tabcolsep}{7pt}
\begin{tabular}{cccccc} \hline\hline
Model ~&~ $m_q$~ &~ $m_c$  ~& $m_b$  ~&~ $\beta_{qc}$ ~&~ $\beta_{qb}$ \\
\hline
Linear ~&~ 0.22~&~ 1.8 ~&~ 5.2 ~&~ 0.4679 ~&~ 0.5266  \\
HO ~&~ 0.25 ~&~ 1.8 ~&~ 5.2 ~&~ 0.4216 ~&~ 0.4960  \\
\hline\hline
\end{tabular}
\end{table}

\begin{figure}[htbp]
\centering
\includegraphics[height=7cm, width=7cm]{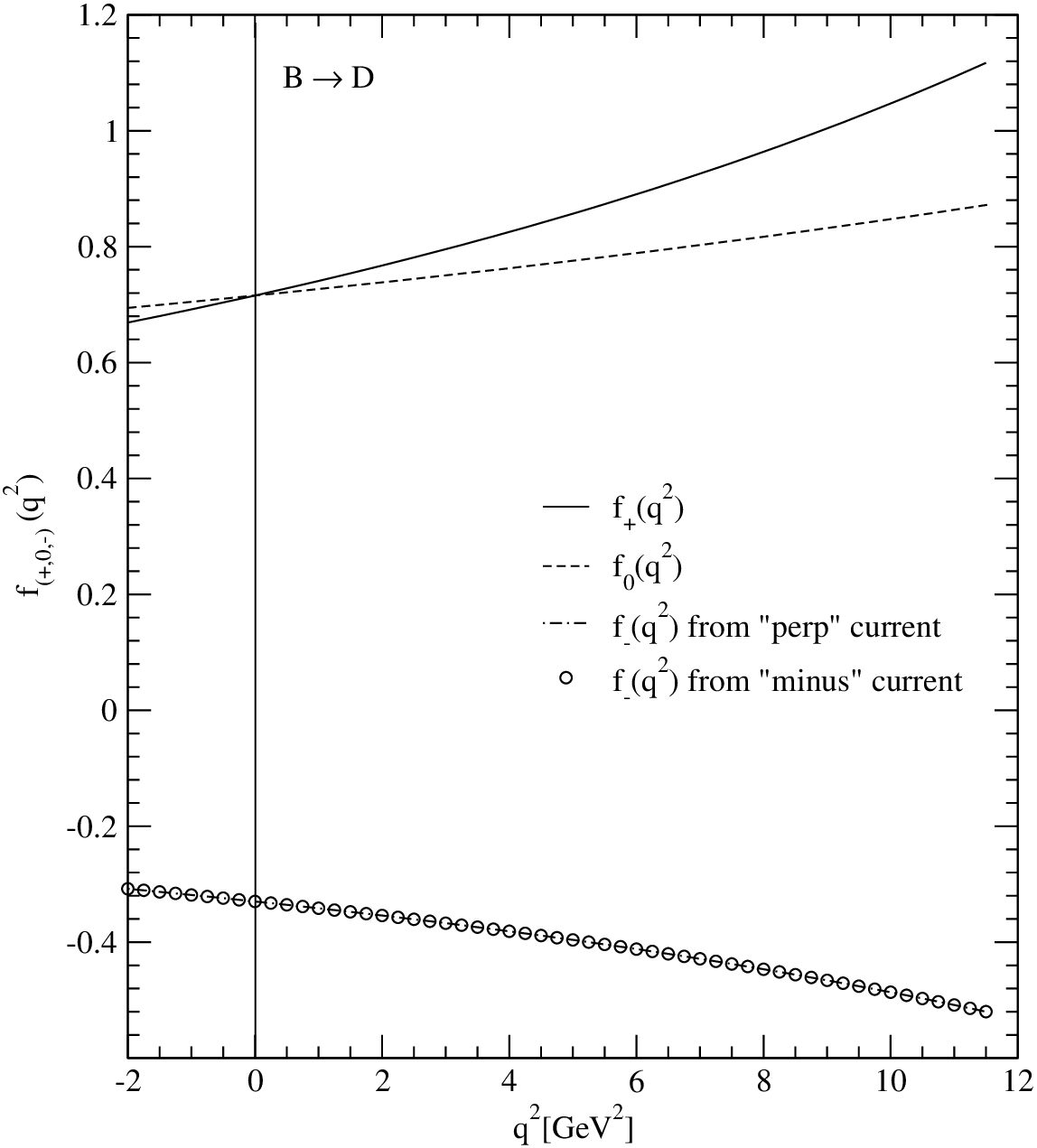}
\caption{\label{fig1} The $q^2$ dependent form factors ($f_+, f_0, f_-$) of the $B\to D\ell\nu_\ell$ decay
for both spacelike and the kinematic timelike regions, $-2\leq q^2\leq (M_B-M_D)^2$ GeV$^2$.}
\end{figure}

In Fig.~\ref{fig1}, we show the $q^2$ dependences of $f_+(q^2)$ (solid line), $f_0(q^2)$ (dashed line),
and $f_-(q^2)$ for $B\to D\ell\nu_\ell$ decay obtained from  Eqs.~(\ref{eq30})-(\ref{eq32}) with the linear potential parameters.
As one can see, our result for $f_-(q^2)$ (dot-dashed line) obtained from $(J^+, {\bf J}_\perp)$ (see Eq.~(\ref{eq31})) shows a complete
agreement with $f_-(q^2)$ (circle) obtained from $(J^+, J^-)$ (see Eq.~(\ref{eq32})) subtantiating the self-consistency of our LFQM.
We also should note that the form factors are displayed not only for 
the whole timelike kinematic region [$m^2_\ell \leq q^2\leq (M_B - M_D)^2$] (in unit of GeV$^2$)
but also for the spacelike region ($-2\leq q^2\leq 0$) (in unit of GeV$^2$) to demonstrate the validity of our analytic continuation from
spacelike region to the timelike by changing ${\bf q}^2_\perp$ to $-{\bf q}^2_\perp (= q^2 >0)$ in the form facfors.

\begin{table}
\caption{Form Factors of the $B\to D\ell\nu_\ell$ decay at $q^2=0$ and $q^2=q^2_{\rm max}$
obtained from the linear (HO) potential parameters.}
\label{t2}
\renewcommand{\arraystretch}{1.2}
\setlength{\tabcolsep}{6pt}
\centering
\begin{tabular}{ccccc} \hline\hline
 $f_+(0)$ & $f_+(q^2_{\rm max})$  & $f_0(q^2_{\rm max})$ & $f_-(0)$ & $f_-(q^2_{\rm max})$ \\
\hline
 0.7157 &  1.1235 & 0.8739 & -0.3298 &  -0.5231\\
(0.6969) &  (1.1209) & (0.8755) & (-0.3190) &  (-0.5142) \\
\hline\hline
\end{tabular}
\end{table}

\begin{table}
\caption{The fitted parameters $b_{+(0)}$ and $c_{+(0)}$ for the parametric form factors in Eq.~(\ref{eq34})
obtained from the linear (HO) potential parameters.}
\label{t3}
\renewcommand{\arraystretch}{1.2}
\setlength{\tabcolsep}{3pt}
\centering
\begin{tabular}{ccccc} \hline\hline
  $f_{+,0}(0)$ &  $b_+$  & $c_+$ & $b_0$ & $c_0$ \\
\hline
0.7157 &  0.955259 & 0.203408 & 0.428416 &  -0.014496\\
(0.6969) &  (1.00776) & (0.245602) & (0.484403) &  (-0.007704)  \\
\hline\hline
\end{tabular}
\end{table}

Our results of the form factors  $(f_\pm, f_0)$ obtained from the linear (HO) potential parameters 
at the maximum recoil ($q^2=0$) and minimum recoil ($q^2=q^2_{\rm max}$) points are 
summarized in Table~\ref{t2}.  Our direct LFQM results for the form factors $f_i (q^2)$ $(i=\pm, 0)$
obtained from Eqs.~(\ref{eq30})-(\ref{eq32}) are well described by the following parametrization~\cite{LCSR1}
\be\label{eq34}
f_{i}(q^2) = \frac{f_{i}(0)}{1 - b_{i} (q^2/M^2_B) + c_{i} (q^2/M^2_B)^2 },
\ee
where the parameters $(b_i, c_i)$ can be obtained from our LFQM results in Eqs.~(\ref{eq30})-(\ref{eq32})
via $b_i=\frac{M^2_B}{f_i(0)} f'_i(0)$ and $c_i = b^2_i - \frac{f''_i(0) M^4_B}{2 f_i(0)}$.
The fitted parameters $(b_i, c_i)$ for $(f_+, f_0)$ are also summarized in Table~\ref{t3} and
those for  $f_-$ are obtained as
$b_-=0.970071\;(1.00817)$ and $c_-=0.200821\;(0.2384)$ for the liner (HO) parameters, respectively.
We should note that our direct
LFQM results and the ones obtained from Eq.~(\ref{eq34}) are in excellent agreement with each other within 0.1$\%$ error.

\begin{figure}
\centering
\includegraphics[height=7cm, width=7cm]{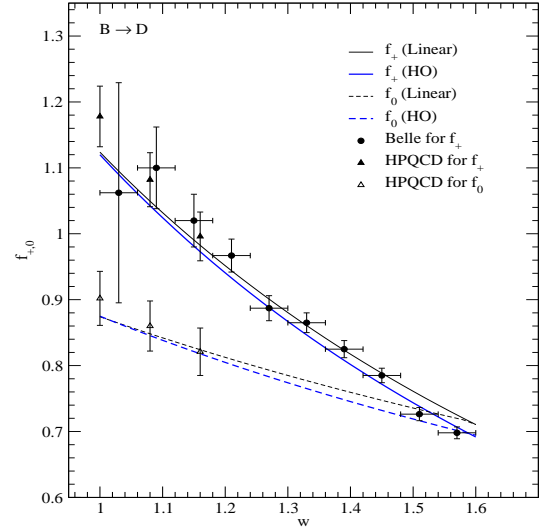}
\caption{\label{fig2} The recoil variable $w$ dependent form factors  $(f_+, f_0)$ of $B\to D\ell\nu_\ell$
obtained from the linear and HO potential parameters, and the result of the combined fit to  
experimental~\cite{Belle16} and lattice QCD (HPQCD)~\cite{Lat1} data.}
\end{figure}

In Fig.~\ref{fig2}, we show the recoil variable $w$ dependent form factor $f_+(w)$ (solid line) and $f_0(w)$ (dashed line) 
obtained from both linear (black lines) and HO (blue line)  potential parameters and compare them with the
data from the Belle experiment~\cite{Belle16} and the lattice QCD (HPQCD)~\cite{Lat1}.
Our results are overall in good agreement with those from ~\cite{Belle16,Lat1}.

Of special interest, while our results for $f_+(w)$ and $f_0(w)$ obtained from the linear potential parameters  (black line)
are somewhat different 
from those obtained from the HO potential parameters (blue lines) at the maximum recoil
point (i.e. $w\simeq 1.6$), both potential parameters give almost the same results at the zero recoil point (i.e. $w=1$).
This is related with the the heavy-quark symmetry (HQS), i.e. in the infinite quark mass limit, the heavy-to-heavy transition form factors
between two pseudoscalar mesons such as $B\to D\ell\nu_\ell$ decay are reduced to single universal Isgur-Wise (IW)
function~\cite{IW1,IW2} ${\cal G}(w)=\frac{2\sqrt{M_B M_D}}{M_B + M_D}f_+(w)$, which should in principle 
satisfy the following normalization ${\cal G}(1)=1$ in the exact HQS limit.
Our LFQM results of  ${\cal G}(1)=0.988\; (0.984)$ obtained from the linear (HO) parameters 
are in good agreement with the exact HQS limit within $2\%$ errors. Our results also 
should be compared with other theoretical
predictions such as  ${\cal G}(1) =1.035 (40)$~\cite{Lat1}, ${\cal G}(1) =1.0541 (83)$~\cite{Lat2}, and ${\cal G}(1)=1.033 (95)$~\cite{Lat3} 
from the lattice QCD and  ${\cal G}(1)=0.981^{+0.045}_{-0.048}$ from the QCD sum rules~\cite{Yi18}.

\begin{figure}
\centering
\includegraphics[height=7cm, width=7cm]{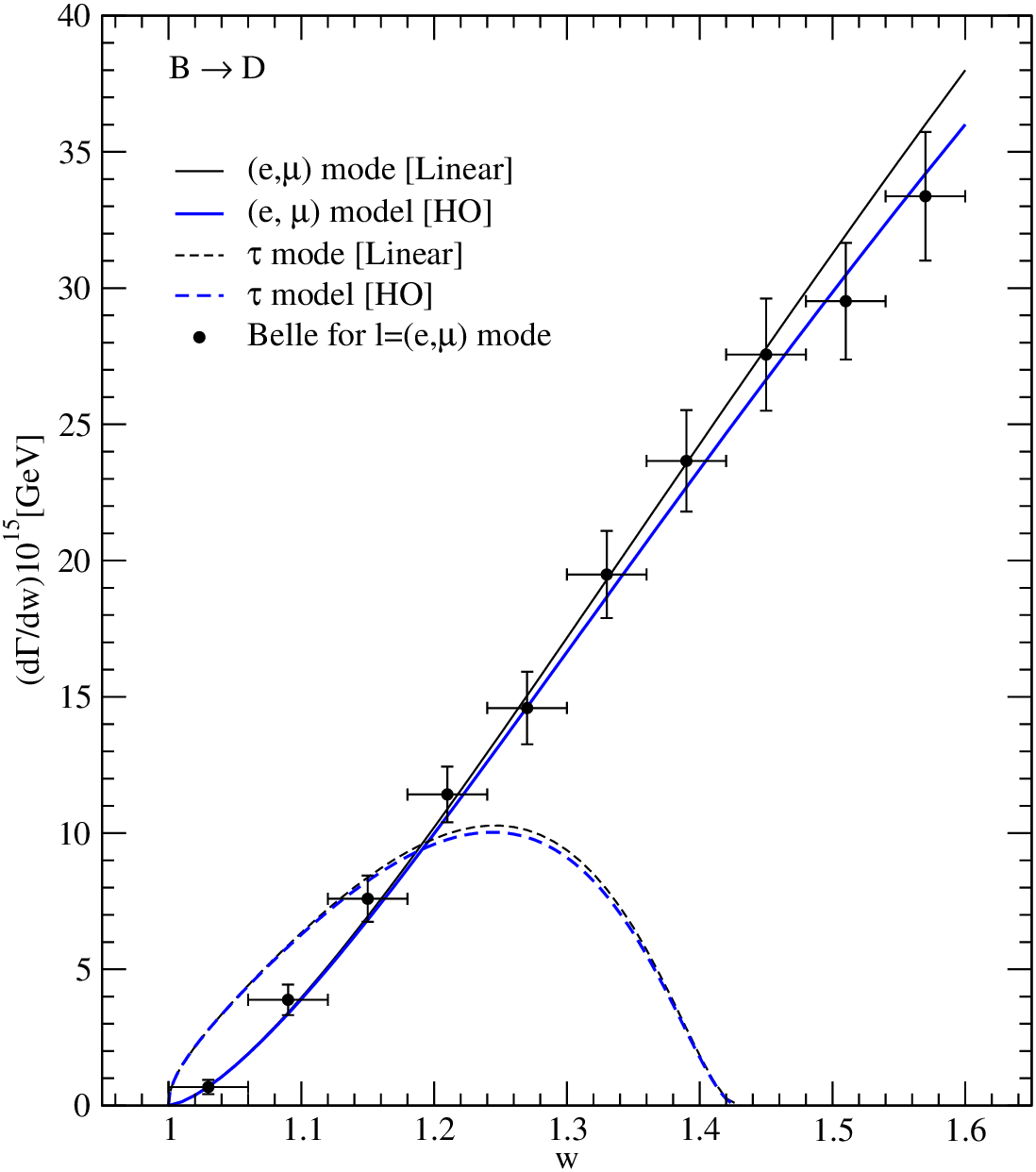}
\caption{\label{fig3} Differential decay width of $B\to D\ell\nu_\ell$ ($\ell=e,\mu,\tau$) compared with the
 experimental data~\cite{Belle16} measured from the light leptonic decay mode.}
\end{figure}

\begin{table*}
\caption{Our LFQM predictions on the branching ratios (in $\%$) for $B\to D\ell\nu_\ell$ ($\ell=e,\mu,\tau$) decays compared with
the results from other theoretical predictions~\cite{LCSR1,HQET} and PDG~\cite{PDG}. $\ell'=e,\mu$.}
\label{t4}
\renewcommand{\arraystretch}{1.2}
\setlength{\tabcolsep}{12pt}
\centering
\begin{tabular}{cccccc} \hline\hline
 Channel &  Linear &HO & LCSR~\cite{LCSR1} & HQET~\cite{HQET} & PDG~\cite{PDG}  \\
\hline
$B^0\to D^-\ell'\nu_{\ell'}$ &  $2.34\pm 0.18$ & $2.25\pm 0.17$ & $2.086^{+0.230}_{-0.232}$ & $-$ &  $2.19\pm 0.12$\\
$B^0\to D^-\tau\nu_\tau$ &  $0.66\pm 0.05$ & $0.64\pm 0.05$ & $0.666^{+0.058}_{-0.057}$ & $0.64\pm 0.05$ &  $1.03\pm 0.22$\\
$B^+\to {\bar D^0}\ell'\nu_{\ell'}$ &  $2.53\pm 0.19$ & $2.44\pm 0.19$ & $2.260^{+0.249}_{-0.251}$ & $-$ &  $2.27\pm 0.11$\\
$B^+\to {\bar D^0}\tau\nu_\tau$ &  $0.72\pm 0.05$ & $0.70\pm 0.05$ & $0.724^{+0.063}_{-0.062}$ & $0.66\pm 0.05$ &  $0.77\pm 0.25$\\
\hline\hline
\end{tabular}
\end{table*}

In Fig.~\ref{fig3}, we show our results for the differential width of  $B\to D\ell\nu_\ell$ ($\ell=e,\mu,\tau$) decay
obtained from both linear (black lines) and HO (blue lines) parameters. The solid lines represent our results for the light 
 ($e, \mu$) decay modes compared with the experimental data from Belle~\cite{Belle16}.
 The dashed lines represent our results for the semitauonic $B\to D\tau\nu_\tau$ decay.
We summarize our LFQM predictions on the branching ratios for $B\to D\ell\nu_\ell$ decays obtained from both linear and HO potential parameters
in Table~\ref{t4} and compare ours with
the results from PDG~\cite{PDG} and other theoretical predictions such as LCSR~\cite{LCSR1}
and heavy quark effective theory (HQET)~\cite{HQET}.
For the numerical calculations of the branching ratios,
we use the CKM matrix element $|V_{cb}|=(40.5\pm 1.5)\times 10^{-3}$,
the PDG values~\cite{PDG} of the lepton ($e,\mu, \tau$) and hadron $(B, D)$ masses 
together with the lifetimes of $(B^0, B^{\pm}$).
As one can see from Table~\ref{t4}, our results obtained from the linear parameters are slightly larger than those obtained
from the HO parameters. Our predictions for three decay modes such as $B^0\to D^-\ell'\nu_{\ell'}$, 
$B^+\to {\bar D^0}\ell'\nu_{\ell'}$, and $B^+\to {\bar D^0}\tau\nu_\tau$ also agree with other theoretical results~\cite{LCSR1,HQET}
as well as PDG values~\cite{PDG} within the errors.
For the semitauonic $B^0\to D^-\tau\nu_\tau$ decay, while three theoretical predictions agree with each other, those theoretical
predictions are smaller than the data from PDG.
From the results given in Table~\ref{t4}, our predictions for the ratio
${\cal R}(D)=\frac{{\rm Br}(B\to D\tau\nu_\tau)}{{\rm Br}(B\to D\ell'\nu_{\ell'})} (\ell'=e,\mu)$ are as follows
\be
{\cal R}(D) = 0.284^{+0.046}_{-0.039} \left[ 0.286^{+0.046}_{-0.040} \right],
\ee
for the linear [HO] potential parameters. Our predictions for the ratio ${\cal R}(D)$ are consistent with other theoretical predictions
such as $0.300(8)$~\cite{Lat1} and $0.299(11)$~\cite{Lat2}  from the LQCD and $0.320^{+0.018}_{-0.021}$~\cite{LCSR1}
within the errors. While our results are quite smaller than the experimental 
values,  ${\cal R}^{\rm exp}(D)=0.440(58)(42)$ from BaBar~\cite{BaB12,BaB13} and
${\cal R}^{\rm exp}(D)=0.375(64)(26)$ from Belle~\cite{Belle15}, we also take note of 
 a new preliminary result ${\cal R}^{\rm exp}(D)=0.307(37)(16)$~\cite{Belle19} reported from the 
Belle collaboration, which is consistent with the SM at the $1.2\sigma$ level.

\section{Summary and Discussion}
\label{sec:V}
In this work, we discussed the self-consistence description on the decay constant $f_{\cal P}$ of a 
pseudoscalar (${\cal P}$) meson and the weak
form factors $f_+$ and $f_-$ (or $f_0$) for the exclusive
semileptonic $B\to D\ell\nu_\ell$ $(\ell=e, \mu,\tau)$ decays in the standard
LFQM. It has been a common perception in the LF formulation that while the plus component ($J^+$) of the LF current $J^\mu$ in 
the matrix element can be regarded as  the ``good" current, the perpendicular (${\bf J}_\perp$) and
the minus ($J^-$) components of the current were known as the ``bad" currents since $({\bf J}_\perp, J^-)$
are easily contaminated by the treacherous points
such as the LF zero mode and the off-mass shell instantaneous contributions.

To scrutinize such treacherous points when the usage of ${\bf J}_\perp$ or $J^-$ is unavoidable, 
we  employed the exactly solvable manifestly covariant BS model  using the multipole type of $q{\bar q}$ bound 
state vertex function. Carrying out the LF calculations for $f_{\cal P}$ and $f_{\pm}(q^2)$
in the BS model, we found that $f_{\cal P}$ and $f_{-}(q^2)$ obtained from the so called ``bad" components of the current
receive the zero-mode contributions as well as the instantaneous ones.
We then linked the covariant BS model to the standard LFQM 
following the same universal correspondence Eq.~(\ref{eq13})
between the two models that we found in our previous analysis of the twist-2 and twist-3 DAs of pseudoscalar and
vector mesons~\cite{CJ14,CJ15,CJ17} and replaced the LF vertex function in the BS model with the more
phenomenologically accessible Gaussian wave function provided by the LFQM analysis of meson mass spectra~\cite{CJ99,CJ99PLB}.
As in the previous analysis~\cite{CJ14,CJ15,CJ17}, it is  striking to observe that the zero mode and the instantaneous contribution
present in the BS model become absent in the LFQM. In other words, 
our LFQM results of the decay constant $f_{\cal P}$ and the TFFs $f_{\pm}(q^2)$ are shown to be independent of the components of the
current without involving any of those treacherous contributions.

We then apply our current independent form factors $f_{\pm}(q^2)$ for the 
self-consistent analysis of $B\to D\ell\nu_\ell$ ($\ell=e,\mu,\tau$) decay
using our LFQM constrained by the variational principle for the QCD-motivated effective Hamiltonian with the linear (or HO) plus
Coulomb interaction~\cite{CJ09,CJ99,CJ99PLB,Choi07}.
The form factors $f_{\pm}(q^2)$ are obtained in the $q^+=0$ frame ($q^2=-{\bf q}^2_\perp <0$) and then analytically continued to the
timelike region by changing ${\bf q}^2_\perp$ to $-q^2$ in the form factors. 
We obtain ${\rm Br}(B\to D\ell\nu_\ell)$ for both neutral and charged $B$ mesons and compare with the experimental data as well as
other theoretical model predictions. Our results for ${\rm Br}(B\to D\ell\nu_\ell)$ show reasonable agreement with the data except
for the semitauonic $B^0\to D^-\tau\nu_\tau$ decay.
Our results for the ratio ${\cal R}(D)$ are consistent with other theoretical predictions as well as the new preliminary result from 
the Belle collaboration~\cite{Belle19} although the previous data from BaBar~\cite{BaB12,BaB13} and Belle~\cite{Belle15}
show quite larger values than our predictions.

\section*{Acknowledgments}
This work was supported by the National Research Foundation of Korea (NRF) 
under Grant No. NRF- 2020R1F1A1067990.

\appendix*
\section{Explicit forms for $f_+(q^2)$ and $f_{-}(q^2)$}
 The on-shell contributions of the trace terms  in Eqs.~(\ref{eq30})-(\ref{eq32}) are given by
 \bea\label{ap:1}
 {\cal T}^+_{\rm on} &=& \frac{4 P^+_1}{{\bar x}} ({\bf k}_\perp\cdot{\bf k'}_\perp
  + {\cal A}_1{\cal A}_2 ),
\nonumber\\
{\cal T}^\perp_{\rm on}
&=& \frac{-2 {\bf k}_\perp}{x {\bar x}} 
\biggl [
 2{\bf k}_\perp\cdot{\bf k'}_\perp + {\bar x} ({\bf q}^2_\perp + m^2_1 + m^2_2)  + 2x^2 m^2_{q}
 \nonumber\\
 &&
 + 2x {\bar x} (m_1m_{q} + m_2m_{q} - m_1m_2) 
 \biggr]
 -\frac{2{\bf q}_\perp}{x {\bar x}} ( {\bf k}^2_\perp + {\cal A}^2_1),
\nonumber\\
{\cal T}^-_{\rm on}
&=& \frac{4}{x^2 {\bar x} P^+} 
\biggl [
{\bar x} (m_1 {\cal A}_1 + {\bf k}^2_\perp) [m^2_2 + ({\bf k}_\perp + {\bf q}_\perp)^2]
\nonumber\\
&&
+ x^2 {\bar x} M^2_0 ({\bf k}^2_\perp + {\bf k}_\perp\cdot{\bf q}_\perp)
+ x^2 m_1 m_2 (m^2_{q} + {\bf k}^2_\perp) 
\nonumber\\
&&+ x {\bar x} m_2 m_{q} (m^2_1 + {\bf k}^2_\perp)
\biggr],
 \eea
 where ${\bar x} =1-x$.
The final standard LFQM results for $f_+(q^2)$ and $f_-(q^2)$  are given by
\bea\label{ap:2}
[ f^{(+)}_{+}]^{\rm SLF}_{\rm on} &=& \int^{1}_{0}dx\int \frac{d^{2}{\bf k}_{\perp}}{16\pi^3}
\frac{\phi_{1}(x,{\bf k}_{\perp})}{\sqrt{ {\cal A}_{1}^{2} + {\bf k}^{2}_{\perp}}}
\frac{\phi_{2}(x,{\bf k}'_{\perp})}{\sqrt{ {\cal A}_{2}^{2}+ {\bf k}^{\prime 2}_{\perp}}}
\nonumber\\
&&\times 
( {\cal A}_{1}{\cal A}_{2}+{\bf k}_{\perp}\cdot{\bf k'}_{\perp} ),
\eea
 \bea\label{ap:3}
[ f^{(\perp)}_-]^{\rm SLF}_{\rm on} &=& \int^1_0 {\bar x} dx
  \int \frac{ d^2{\bf k}_\perp } { 16\pi^3 }
  \frac{ \phi_1 (x, {\bf k}_\perp) } {\sqrt{ {\cal A}^2_1 + {\bf k}^2_\perp }}
  \frac{ \phi_2 (x, {\bf k'}_\perp) } {\sqrt{ {\cal A}^2_2 + {\bf k}^{\prime 2}_\perp }}
  \nonumber\\
  &&\times
  \biggl[ - {\bar x} M^2_0 + (m_2 - m_q){\cal A}_1 - m_q (m_1 - m_q)
  \nonumber\\
  &&+ \frac{ {\bf k}_\perp\cdot{\bf q}_\perp }{q^2} [ M^2_0 + M'^2_0
  - 2(m_1 - m_q) (m_2 - m_q) ]
  \biggr],
  \nonumber\\
   \eea
   and 
  \bea\label{ap:4}
[f^{(-)}_{-}]^{\rm SLF}_{\rm on} &=&
 \int^1_0\frac{dx}{x^2} \int\frac{d^2{\bf k}_\perp}{16\pi^3} 
 \frac{\phi_{1}(x,{\bf k}_{\perp})}{\sqrt{ {\cal A}_{1}^{2} + {\bf k}^{2}_{\perp}}}
\frac{\phi_{2}(x,{\bf k}'_{\perp})}{\sqrt{ {\cal A}_{2}^{2}+ {\bf k}^{\prime 2}_{\perp}}}
\nonumber\\
&&\times
\biggl\{ 
a_0
  \biggl[ x^2 {\bar x} M^2_0 ({\bf k}^2_\perp + {\bf k}_\perp\cdot{\bf q}_\perp)
  \nonumber\\
  &&
  + {\bar x} (m_1 {\cal A}_1 + {\bf k}^2_\perp) [m^2_2 + ({\bf k}_\perp + {\bf q}_\perp)^2]
\nonumber\\
&&
+ x^2 m_1 m_2 (m^2_{q} + {\bf k}^2_\perp) 
+ x {\bar x}  m_2 m_{q} (m^2_1 + {\bf k}^2_\perp)
\biggr]
\nonumber\\
&&
- x^2  b_0 ( {\bf k}_\perp\cdot{\bf k'}_\perp + {\cal A}_1{\cal A}_2 )
\biggl\},
 \eea
 where $a_0 = \frac{2}{M^2_0 - M'^2_0 -{\bf q}_\perp^2}$
 and $b_0 = \frac{M^2_0 + M'^2_0 + {\bf q}_\perp^2}{M^2_0 - M'^2_0 -{\bf q}_\perp^2}$.


\end{document}